\documentclass{iaus}
\usepackage{graphicx}
\usepackage{amssymb}
\usepackage{amsmath}
\usepackage{bm}
\usepackage{upmath}
\usepackage{color}

\newcommand{\be}{\begin{equation}}
\newcommand{\ee}{\end{equation}}
\newcommand{\ba}{\begin{eqnarray}}
\newcommand{\ea}{\end{eqnarray}}
\renewcommand{\l}{\label}
\newcommand{\f}{\frac}

\renewcommand{\a}{\alpha}
\renewcommand{\b}{\beta}
\renewcommand{\d}{\delta}

\renewcommand{\p}{\partial}
\renewcommand{\le}{\left}
\renewcommand{\r}{\right}
\newcommand{\aj}{{\it Astron. J. (USA)}}

\newcommand{\pla}{{{\it Phys. Lett.}  A}}
\newcommand{\plb}{{{\it Phys. Lett.}  B}}

\newcommand{\prd}{{{\it Phys. Rev.} D}}

\newcommand{\prep}{{\it Phys. Reports}}

\newcommand{\cmda}{{\it Cel. Mech. Dyn. Astron.}}

\newcommand{\aap}{{\it Astron. Astrophys.}}
\newcommand{\cqg}{{\it Class. Quant. Grav.}}

\title[Beyond the Standard IAU Framework ] 
{Beyond the Standard IAU Framework }

\author[Sergei Kopeikin]   
{Sergei Kopeikin}
\affiliation{Department of Physics \& Astronomy, University of Missouri-Columbia,\\ Columbia, MO 65211, USA}

\pubyear{2009}
\volume{261}  
\pagerange{1--9}
\jname{Relativity in Fundamental Astronomy: \\Dynamics, Reference Frames, and Data Analysis}
\editors{S. Klioner, P.K. Seidelmann \& M. Soffel, eds.}
\begin{document}

\maketitle

\begin{abstract}
We discuss three conceivable scenarios of extension and/or modification of the IAU relativistic resolutions on time scales and spatial coordinates beyond the Standard IAU Framework. These scenarios include: (1) the formalism of the monopole and dipole moment transformations of the metric tensor replacing the scale transformations of time and space coordinates; (2) implementing the parameterized post-Newtonian formalism with two PPN parameters - ${\beta}$ and $\gamma$; (3) embedding the post-Newtonian barycentric reference system to the Friedman-Robertson-Walker cosmological model.
\end{abstract}

\firstsection 
              \section{Introduction}
Fundamental scientific program of detection of gravitational waves by the space interferometric gravitational wave detectors
like LISA is a driving motivation for further systematic developing of relativistic theory of reference frames in the solar system and beyond. LISA will detect gravitational wave sources from all directions in the sky. These sources will include thousands of compact binary systems containing neutron stars, black holes, and white dwarfs in our own Galaxy, and merging super-massive black holes in distant galaxies. However, the detection and proper interpretation of the gravitational waves can be achieved only under the condition that all coordinate-dependent effects are completely understood and subtracted from the signal. This is especially important for observation of gravitational waves from very distant sources located at cosmological distances because the Hubble expansion of our universe affect propagation of the waves.

New generation of microarcsecond astrometry satellites: SIM
and a cornerstone mission of ESA - Gaia, requires a novel approach for an unambiguous
interpretation of astrometric data obtained from the on-board
optical instruments. SIM and Gaia complement one another. Both SIM and Gaia will approach the accuracy of 1 $\mu$as.
Gaia will observe all stars ($\sim 10^9$) between magnitude 6 and 20.  The
accuracy of Gaia is about 5 $\mu$as for the optimal stars (magnitude
between 6 and 13).  SIM is going to observe \~10000 stars with magnitude
up to 20.  The accuracy of SIM is expected to be a few $\mu$as and can be
reached for any object brighter than about 20 provided that sufficient
observing time is allocated for that object. At this level the problem of propagation of light rays must be treated with taking into account relativistic effects generated by non-static part of the gravitational field of the solar system and binary stars (\cite{kopg}). Astrometric resolution in 1 $\mu$as forces us to
change the classic treatment of parallax, aberration, and proper motion
of stars by switching to a more precise definition of reference frames on a curved space-time manifold (\cite{ks,km,kkp}). Advanced practical realization of an inertial reference frame is required for unambiguous physical interpretation of the gravitomagnetic precession of orbits of LAGEOS satellites (\cite{ciuf,ries}) and GP-B gyroscope, which is measured relative to a binary radio star IM Pegasi that has large annual parallax and proper motion (\cite{bartel}).

Recent breakthroughs in technology of drag-free satellites, clocks, lasers, optical and radio interferometers and new demands of experimental gravitational physics (\cite{leh,hlnt}) make it necessary to incorporate the parameterized post-Newtonian formalism (\cite{will}) to the procedure of construction of relativistic local frames around Earth and other bodies of the solar system (\cite{klis1,kv}).
The domain of applicability of the IAU relativistic theory of
reference frames (\cite{iau2}) should be also extended outside the boundaries of the solar system (\cite{kopg}).

In what follows, Latin indices takes values 1,2,3, and the Greek ones run from 0 to 3. Repeated indices indicate the Einstein summation rule. The unit matrix is denoted $\delta_{ij}={\rm diag}(1,1,1)$ and the fully anti-symmetric symbol $\epsilon_{ijk}$ is subject to $\epsilon_{123}=1$. The Minkowski metric is $\eta_{\a\b}={\rm diag}(-1,1,1,1)$. Greek indices are raised and lowered with the Minkowski metric, Latin indices are raised and lowered with the unit matrix. Bold italic letters denote spatial vectors. Dot and cross between two spatial vectors denote the Euclidean scalar and vector products respectively. Partial derivative with respect to spatial coordinates $x^i$ are denoted as $\p/\p x^i$ or ${\vec\nabla}$.

\section{Standard IAU Framework}
New relativistic resolutions on
reference frames and time scales in the solar system were
adopted by the 24-th General Assembly of the IAU in 2000 (\cite{iau2,soff}). The resolutions are based on the first post-Newtonian
approximation of general relativity. They abandoned the Newtonian paradigm of
space and time and required corresponding change in the conceptual basis and terminology of the fundamental astronomy (\cite{iau06}).

Barycentric Celestial Reference System (BCRS), $x^\a=(ct,{\bm x})$,
is defined in terms of a metric tensor $g_{\a\b}$ with components
\begin{eqnarray}
g_{00} &=& - 1 + \f{2 w}{c^2} - \f{2w^2}{c^4} + O(c^{-5})\;,
\label{5} \\
g_{0i} &=& - \f{4 w^i}{c^3} + O(c^{-5})\;,\label{6}
 \\
g_{ij} &=& \delta_{ij}
\left( 1 + \f{2w}{c^2} \right) + O(c^{-4})\; .
\label{7}
\end{eqnarray}
Here, the post-Newtonian gravitational potential $w$ generalizes the
Newtonian potential, and $w^i$ is a vector potential related
to the gravitomagnetic effects (\cite{cw}). These potentials are defined by solving the
field equations
\begin{eqnarray}
\label{eq1}
\Box w&=&-4\pi G\sigma\;,\\
\label{eq2}
\Box w^i&=&-4\pi G\sigma^i\;,
\end{eqnarray}
where $\Box\equiv -c^{-2}\partial^2/\partial t^2+\nabla^2$ is the wave operator,
$\sigma = c^{-2}(T^{00} + T^{ss}),$ $
\sigma^i = c^{-1}T^{0i}$, and
$T^{\mu\nu}$ are the components of the stress-energy tensor of the solar system bodies, $T^{ss}= T^{11} + T^{22} + T^{33}$.

Equations (\ref{eq1}), (\ref{eq2})
are solved by iterations
\begin{eqnarray}
w(t,{\bm x}) &=& G \int
\f{\sigma(t, {\bm x}')d^3 x'}{\vert {\bm x} - {\bm x}' \vert}
 + \f{G }{2c^2}  \f{\partial^2}{\partial t^2}
\int d^3 x'  \sigma(t,{\bm x}') \vert {\bm x} - {\bm x}' \vert +O(c^{-4})\; ,
\label{8}\\\label{9}
w^i(t,{\bm x}) &=& G \int\f{\sigma^i (t,{\bm x}') d^3 x'}{
\vert{\bm x} - {\bm x}' \vert } +O(c^{-2})\;,
\end{eqnarray}
which are to be substituted to the metric tensor (\ref{5})--(\ref{7}). Each of the potentials, $w$ and $w^i$, can be linearly decomposed in two parts
\ba\l{eq3}
w&=&w_E+{\bar w}\;,\\\l{eq4}
w^i&=&w_E^i+{\bar w}^i\;,
\ea
where $w_E$ and $w_E^i$ are BCRS potentials depending on the distribution of mass and current only inside the Earth, and  ${\bar w}_E$ and ${\bar w}_E^i$ are gravitational potentials of external bodies.

Geocentric Celestial Reference System (GCRS) is denoted $X^\a=(cT, {\bm X})$.
GCRS is defined in terms of the metric tensor $G_{\a\b}$ with components
\begin{eqnarray}
G_{00} &=& - 1 + \f{2 W}{c^2} - \f{2W^2 }{ c^4} + O(c^{-5})\;,
\label{11} \\\label{12}
G_{0i} &=& - \f{4 W^i }{c^3} + O(c^{-5})\;,
 \\
G_{ij} &=& \delta_{ij}
\left( 1 + \f{2 W }{ c^2} \right) + O(c^{-4})\; .
\label{13}
\end{eqnarray}
Here $W = W(T,{\bm X})$ is the post-Newtonian gravitational potential and $W^i(T,{\bm X})$ is a vector-potential both expressed in the geocentric coordinates. They satisfy to the same type of the wave equations (\ref{eq1}), (\ref{eq2}).

The geocentric potentials are split into two parts: potentials
$W_E$ and $W_E^i$ arising from the gravitational field of the Earth
and external parts associated with tidal and inertial effects. IAU resolutions implied that the external parts must vanish at the
geocenter and admit an expansion in powers of ${\bm X}$ (\cite{iau2,soff})
\begin{eqnarray}
W(T,{\bm X}) &=& W_E(T,{\bm X})
+ W_{\rm kin}(T,{\bm X})+W_{\rm dyn}(T,{\bm X})
\;,
\label{14}\\
\label{15}
W^i(T,{\bm X}) &=& W^i_E(T,{\bm X})
+ W^i_{\rm kin}(T,{\bm X})+W^i_{\rm dyn}(T,{\bm X})
\;.
\end{eqnarray}
Geopotentials  $W_E$ and $W^i_E$
are defined in the same way as $w_E$ and $w_E^i$ (see equations (\ref{8})--(\ref{9})) but with quantities $\sigma$ and $\sigma^i$
calculated in the GCRS. $W_{\rm kin}$ and $W_{\rm kin}^i$ are kinematic contributions that are linear in spatial coordinates ${\bm X}$
\begin{equation}\label{16}
W_{\rm kin}= Q_i X^i\;,\qquad\qquad
W^i_{\rm kin}= \f14\;c^2
\varepsilon_{ipq} (\Omega^p - \Omega^p_{\rm prec})\;X^q\;,
\end{equation} where
$Q_i$ characterizes the minute deviation of the actual world line of
the geocenter from that of a fiducial test particle being in a free fall in the
external gravitational field of the solar system bodies (\cite{kop88})
\begin{equation}\label{17}
Q_i=\partial_i {\bar w}({\bm x}_E)-a_E^i+O(c^{-2})\;.
\end{equation}
Here
$a_E^i={dv^i_E/dt}$ is the
barycentric acceleration of the geocenter. Function
$\Omega^a_{\rm prec}$ describes the relativistic precession
of dynamically non-rotating spatial axes of GCRS with respect to reference quasars
\begin{eqnarray}\label{18}
\Omega_{\rm prec}^i =
\f{1}{c^2}\,
\varepsilon_{ijk}\,
\left(
-\f32\,v^j_E\,\partial_k {\bar w}({\bm x}_E)
+2\,\partial_k {\bar w}^j({\bm x}_E)
-\f12\,v^j_E\,Q^k
\right).
\end{eqnarray}
The three terms on the right-hand side of this equation
represent the geodetic, Lense-Thirring, and Thomas precessions,
respectively (\cite{kop88,iau2}).

Potentials $W_{\rm dyn}$ and $W^i_{\rm dyn}$ are generalizations
of the Newtonian tidal potential.
For example,
\begin{equation}\label{19}
W_{\rm dyn}(T,{\bm X}) =
{\bar w}({\bm x}_E + {\bm X}) - {\bar w}({\bm x}_E) - X^i \p_i{\bar w}({\bm x}_E)+O(c^{-2}).
\end{equation}
It is easy to check out that a Taylor expansion of ${\bar w}({\bm x}_E + {\bm X})$ around the point ${\bm x}_E$ yields $W_{\rm dyn}(T,{\bm X})$ in the form of a polynomial starting from the quadratic with respect to ${\bm X}$ terms.
We also note that the local gravitational potentials $W_E$ and $W_E^i$ of the
Earth are related to the barycentric gravitational potentials $w_E$ and
$w^i_E$ by the post-Newtonian transformations (\cite{bk,iau2}).

\section{IAU Scaling Rules and the Metric Tensor}

The coordinate transformations between the BCRS and GCRS are found by matching the BCRS and GCRS metric tensors in the vicinity of the world line of the Earth by making use of their tensor properties. The transformations are written as (\cite{kop88,iau2})
\begin{eqnarray}
\label{22}
T&=&t - \f{1}{ c^2} \left[ A + {\bm v}_E\cdot{\bm r}_E \right]
+ \f{1}{ c^4} \left[ B + B^ir_E^i +
B^{ij}r_E^ir_E^j + C(t,{\bm x}) \right] +O(c^{-5}),
\\\label{23}
X^i&=&
r^i_E+\frac 1{c^2}
\left[\frac 12 v_E^i {\bm v}_E\cdot{\bm r}_E + {\bar w}({\bm x}_E) r^i_E
+ r_E^i {\bm a}_E\cdot{\bm r}_E-\frac 12 a_E^i r_E^2
\right]+O(c^{-4}),
\end{eqnarray}
where ${\bm r}_E={\bm x}-{\bm x}_E$, and functions $A, B, B^i, B^{ij}, C(t,{\bm x})$ are
\def\xe{({\bm x}_E)}
\begin{eqnarray}
\f{dA}{dt}&=&\f12\,v_E^2+{\bar w}\xe,\label{24}
\\
\f{dB}{ dt}&=&-\f{1}{8}\,v_E^4-\f{3}{ 2}\,v_E^2\,{\bar w}\xe
+4\,v_E^i\,{\bar w}^i+\f{1}{2}\,{\bar w}^2\xe,\label{25}
\\\label{26}
B^i&=&-\f{1}{2}\,v_E^2\,v_E^i+4\,{\bar w}^i\xe-3\,v_E^i\,{\bar w}\xe,
\\\label{27}
B^{ij} &=& -v_E^{i} Q_{j}+
2 \p_j {\bar w}^i({\bm x}_E)
-v_E^{i} \p_j {\bar w}\xe+\f{1}{ 2}
\,\delta^{ij} \dot{{\bar w}}({\bm x}_E),
\\\label{28}
C(t,{\bm x})&=&-\f{1}{10}\,r_E^2\,({\dot{\bm a}_E}\cdot{\bm r}_E)\, .
\end{eqnarray}
Here again $x_E^i, v_E^i$, and $a_E^i$ are the barycentric position,
velocity and acceleration vectors of the Earth, the dot stands for the
total
derivative with respect to $t$.
The harmonic gauge
condition does not fix the function $C(t,{\bm x})$ uniquely. However, it is reasonable to fix it in the time transformation for practical reasons (\cite{iau2}).

Earth's orbit in BCRS is almost circular. This makes the right side of equation (\ref{24}) almost constant with small periodic oscillations
\begin{equation}\label{u1}
\f12\,v_E^2+{\bar w}\xe=c^2L_C+(\mbox{periodic terms})\;,
\end{equation}
where the constant $L_C$ and the periodic terms have been calculated with a great precision in (\cite{fuki}). For practical reason of calculation of ephemerides of the solar system bodies, the BCRS time coordinate was re-scaled to remove the constant $L_C$ from the right side of equation (\ref{u1}). The new time scale was called TDB
\begin{equation}\label{u2}
t_{TDB}=t\left(1-L_B\right)\;,
\end{equation}
where a new constant $L_B$ is used for practical purposes instead of $L_C$ in order to take into account the additional linear drift between the geocentric time $T$ and the atomic time on geoid, as explained in (\cite{tmsc,irf,ksei}).
Time re-scaling changes the Newtonian equations of motion. In order to keep the equations of motion invariant scientists doing the ephemerides also re-scaled spatial coordinates and masses of the solar system bodies. These scaling transformations are included to
IAU 2000 resolutions (\cite{iau2}) but they have never been explicitly associated with transformation of the metric tensor. It had led to a long-standing discussion about the units of measurement of time, space, and mass in astronomical measurements (\cite{ksei}).

The scaling of time and space coordinates is associated with a particular choice of the metric tensor corresponding to either TCB or TDB time. In order to see it, we notice that equation (\ref{16}) is a solution of the Laplace equation, which is defined up to an arbitrary function of time $Q$. If one takes it into account, equations (\ref{16}), (\ref{24}) can be re-written as
\begin{eqnarray}\label{u3}
W_{\rm kin}&=&Q+Q_i X^i\;,\\
\label{u4}
\f{dA(t)}{dt}&=&\f12\,v_E^2+{\bar w}\xe-Q\;,
\end{eqnarray}
and, if we chose $Q=c^2 L_C$, it eliminates the secular drift between times $T$ and $t$ without explicit re-scaling of time, which is always measured in SI units. It turns out that Blanchet-Damour (\cite{bdmass}) relativistic definition of mass depends on function Q and is re-scaled in such a way that the Newtonian equations of motion remain invariant. Introduction of $Q$ to function $W$ brings about implicitly re-scales spatial coordinates as well. We conclude that introducing function $Q=c^2 L_C$ to the metric tensor without apparent re-scaling of coordinates and masses might be more preferable for IAU resolutions as it allows us to keep the SI system of units without changing coordinates and masses made ad hoc "by hands".
Similar procedure can be developed for re-scaling the geocentric time $T$ to take into account the linear drift existing between this time and the atomic clocks on geoid (\cite{tmsc,irf}).

\section{Parameterized Coordinate Transformations}
This section discusses how to incorporate the parameterized post-Newtonian (PPN) formalism (\cite{will}) to the
IAU resolutions.  This extends applicability of the resolutions[
to a more general class of gravity theories. Furthermore, it makes
the IAU resolutions fully compatible with JPL equations of motion used for calculation of ephemerides of major
planets, Sun and Moon.

These equations of motion depend on two PPN parameters, ${\beta}$ and $\gamma$ (\cite{sman}) and they
are presently compatible with the IAU resolutions only in the case of $\beta=\gamma=1$.
Rapidly growing precision of optical and radio astronomical
observations as well as calculation of relativistic equations of motion in gravitational wave astronomy urgently
demands to work out a PPN theory of relativistic
transformations between the local and global coordinate systems.

PPN parameters ${\beta}$ and $\gamma$ are characteristics of a scalar field which makes the metric tensor different from general relativity.
In order to extend the IAU 2000 theory of reference frames to the PPN formalism we employ a general class of Brans-Dicke theories (\cite{brd}).
This class is
based on the metric tensor $g_{\alpha\beta}$ and a scalar
field $\phi$ that couples with the metric tensor through
function $\theta(\phi)$. We assume that $\phi$ and $\theta(\phi)$ are analytic functions
which can be expanded in a Taylor series about their background values $\bar{\phi}$ and $\bar{\theta}$.

The parameterized
theory of relativistic reference
frames in the solar system is built in accordance to the same rules as used in the IAU resolutions. The entire procedure is described in our paper (\cite{kv}).
The parameterized coordinate transformations between BCRS and GCRS are found by matching the BCRS and GCRS metric tensors and the scalar field in the vicinity of the world line of the Earth. The transformations have the following form (\cite{kv})
\begin{eqnarray}
\label{22q}\hspace{-1cm}
T&=&t - \f{1}{ c^2} \left[ A + {\bm v}_E\cdot{\bm r}_E \right]
+ \f{1}{ c^4} \left[ B + B^i\,r_E^i +
B^{ij}\,r_E^i\,r_E^j + C(t,{\bm x}) \right] +O(c^{-5}),
\\\label{23q}
X^i&=&
r^i_E+\frac 1{c^2}
\left[\frac 12 v_E^i v_E^jr^j_E +\gamma Qr^i_E+ \gamma{\bar w}({\bm x}_E)r^i_E
+ r_E^i a^j_E r^j_E-\frac 12 a_E^i r_E^2
\right]+O(c^{-4})
\end{eqnarray}
where ${\bm r}_E={\bm x}-{\bm x}_E$, and functions $A(t), B(t), B^i(t), B^{ij}(t), C(t,{\bm x})$ are
\def\xe{({\bm x}_E)}
\begin{eqnarray}
\f{dA}{ dt}&=&\f{1}{2}\,v_E^2+{\bar w}-Q\xe,\label{24q}
\\
\f{dB}{ dt}&=&-\f{1}{8}\,v_E^4-\le(\gamma+\f{1}{ 2}\r)\,v_E^2\,{\bar w}\xe
+2(1+\gamma)\,v_E^i\,{\bar w}^i+\le(\beta-\f{1}{ 2}\r)\,{{\bar w}}^2\xe,\label{25q}
\\\label{26q}
B^i&=&-\f{1}{2}\,v_E^2\,v_E^i+2(1+\gamma)\,{\bar w}^i\xe-(1+2\gamma)\,v_E^i\,{\bar w}\xe,
\\\label{27q}
B^{ij} &=& -v_E^{i} Q_{j}+
(1+\gamma) \p_j {\bar w}^i({\bm x}_E)
-\gamma v_E^{i} \p_j {\bar w}\xe+\f{1}{ 2}
\,\delta^{ij} \dot{{\bar w}}({\bm x}_E),
\\\label{28q}
C(t,{\bm x})&=&-\f{1}{10}\,r_E^2\,({\dot{\bm a}_E}\cdot{\bm r}_E)\, .
\end{eqnarray}
These transformations depends explicitly on the PPN parameters ${\beta}$ and $\gamma$ and the scaling function $Q$, and should be compared with those (\ref{22})-(\ref{28}) adopted in the IAU resolutions.

PPN parameters ${\beta}$ and $\gamma$ have a fundamental physical meaning in the scalar-tensor theory of gravity along with the universal gravitational constant $G$ and the fundamental speed $c$. It means that if the parameterized transformations (\ref{22q})-(\ref{28q}) are adopted by the IAU, the parameters ${\beta}$ and $\gamma$ must be included to the number of the astronomical constants which values must be determined experimentally. The program of the experimental determination of ${\beta}$ and $\gamma$ began long time ago and it makes use of various observational techniques. So far, the experimental values of ${\beta}$ and $\gamma$ are indistinguishable from their general-relativistic values $\beta=1$, $\gamma=1$.

\section{Matching IAU Resolutions with Cosmology}

BCRS assumes that the solar system is isolated and space-time is asymptotically flat. This idealization will not work at some level of accuracy of astronomical observations because the universe is expanding and its space-time is described by the Friedman-Robertson-Walker (FRW) metric tensor having non-zero Riemannian curvature (\cite{mtw}). It may turn out that some, yet unexplained anomalies in the orbital motion of the solar system bodies (\cite{p-conf,krbr,dittus}) are indeed associated with the cosmological expansion. Moreover, astronomical observations of cosmic microwave background radiation and other cosmological effects requires clear understanding of how the solar system is embedded to the cosmological model. Therefore, it is reasonable to incorporate the cosmological expansion of space-time to the IAU 2000 theory of reference frames in the solar system (\cite{kram,ramk}).

Because the universe is not asymptotically-flat the gravitational field of the solar system can not vanish at infinity. Instead, it must match with the cosmological metric tensor. This imposes the cosmological boundary condition. The cosmological model is not unique has a number of free parameters depending on the amount of visible and dark matter, and on the presence of dark energy. We considered a FRW universe driven by a scalar field imitating the dark energy $\phi$ and having a spatial curvature equal to zero (\cite{sfi1,sfi2}). The universe is perturbed by a localized distribution of matter of the solar system. In this model the perturbed metric tensor reads
\begin{equation}
\label{c1}
g_{\alpha\beta}=a^2(\eta)f_{\alpha\beta}\;,\qquad f_{\alpha\beta}=(\eta_{\alpha\beta}+h_{\alpha\beta})\;,
\ee
where perturbation $h_{\alpha\beta}$ of the background FRW metric tensor $\bar g_{\a\b}=a^2\eta_{\a\b}$ is caused by the presence of the solar system, $a(\eta)$ is a scale factor of the universe depending on the, so-called, conformal time $\eta$ related to coordinate time $t$ by simple differential equation $dt=a(\eta)d\eta$. In what follows, a linear combination of the metric perturbations
\be\l{c1a}
\gamma^{\alpha\beta}=h^{\alpha\beta}-\f12\eta^{\alpha\beta}h\;,
\ee
where $h=\eta^{\alpha\beta}h_{\alpha\beta}$, is more convenient for calculations.

We discovered a new cosmological gauge, which has a number of remarkable properties. In case of the background FRW universe with dust equation of state (that is, the background pressure of matter is zero (\cite{sfi1,sfi2}) this gauge is given by (\cite{kram,ramk})
\begin{equation}
\label{c2}
\gamma^{\a\b}{}_{|\b}=2H\varphi \d^\a_0\;,
\end{equation}
where bar denotes a covariant derivative with respect to the background metric $\bar g_{\a\b}$, $\varphi=\phi/a^2$, $H=\dot a/a$ is the Hubble parameter, and the overdot denotes a time derivative with respect to time $\eta$. The gauge (\ref{c2}) generalizes the harmonic gauge of asymptotically-flat space-time for the case of the expanding non-flat background universe.

The gauge (\ref{c2}) drastically simplifies the linearized Einstein equations. Introducing notations $\gamma_{00}\equiv 4w/c^2$, $\gamma_{0i}\equiv -4w^i/c^3$, and $\gamma_{ij}\equiv 4w^{ij}/c^4$, and splitting Einstein's equations in components, yield
\begin{eqnarray}\label{c5}
\Box\chi-2H\p_\eta \chi+\f52H^2\chi&=&-4\pi G\sigma\;,
\\\label{c6}
\Box w-2H\p_\eta w\phantom{oooo[[::::}&=&-4\pi G\sigma-4H^2\chi\;,\\
\label{c7}
\Box w^{i}-2H\p_\eta w^{i}+H^2 w^{i}&=&-4\pi G\sigma^{i}\;,\\
\label{c8}
\Box w^{ij}-2H\p_\eta w^{ij}\phantom{ooooo}&=&-4\pi G T^{ij}\;,
\end{eqnarray}
where $\partial_\eta\equiv\p/\p\eta$,  $\,\Box\equiv -c^{-2}\p^2_\eta+{\bm\nabla}^2$, $\chi\equiv w-\varphi/2$, the Hubble parameter $H=\dot a/a=2/\eta$, densities $\sigma=c^{-2}(T^{00}+T^{ss})$, $\sigma^i=c^{-1}T^{0i}$ with $T^{\alpha\beta}$ being the tensor of energy-momentum of matter of the solar system defined with respect to the metric $f_{\alpha\beta}$. These equations extend the domain of applicability of equations (\ref{eq1}), (\ref{eq2}) of the IAU standard framework (\cite{soff}) to the case of expanding universe.

First equation (\ref{c5}) describes evolution of the scalar field while the second equation (\ref{c6}) describes evolution of the scalar perturbation of the metric tensor. Equation (\ref{c7}) yields evolution of vector perturbations of the metric tensor, and equation (\ref{c8}) describes generation and propagation of gravitational waves by the isolated N-body system. Equations (\ref{c5})--(\ref{c8})
contain all corrections depending on the Hubble parameter and can be solved analytically in terms of generalized retarded solution. Exact Green functions for these equations have been found in (\cite{kram,ramk,pois1}). They revealed that the gravitational perturbations of the isolated system on expanding background depend not only on the value of the source taken on the past null cone but also on the value of the gravitational field inside the past null cone.

Existence of extra terms in the solutions of equations  (\ref{c5})--(\ref{c8}) depending on the Hubble parameter brings about cosmological corrections to the Newtonian law of gravity. For example, the post-Newtonian solutions of equations (\ref{c6}), (\ref{c7}) with a linear correction due to the Hubble expansion are
\ba\l{cor1}
w(\eta,{\bm x}) &=& G \int  \;
\f{\sigma(\eta, {\bm x}')d^3 x'}{\vert {\bm x} - {\bm x}' \vert}
 + \f{G }{2c^2}  \f{\partial^2}{\partial \eta^2}
\int d^3 x' \sigma(\eta,{\bm x}') \vert {\bm x}- {\bm x}' \vert \\\nonumber & -& GH\int d^3 x'\sigma(\eta, {\bm x}') +O\left(c^{-4}\right)+O\left(H^2\right)\; ,
\\\label{cor2}
w^i(\eta,{\bm x}) &=& G \int\; \f{\sigma^i (\eta,{\bm x}') d^3 x'}{
\vert{\bm x} - {\bm x}' \vert } +O\left(c^{-2}\right)+O\left(H^2\right)\;.
\end{eqnarray}\noindent
Matching these solutions with those defined in the BCRS of the IAU 2000 framework in equations (\ref{8}), (\ref{9}) is achieved after expanding all quantities depending on the conformal time $\eta$ in the neighborhood of the present epoch in powers of the Hubble parameter.

Current IAU 2000 paradigm assumes that the asymptotically-flat metric, $f_{\alpha\beta}$, is used for calculation of light propagation and ephemerides of the solar system bodies. It means that the conformal time $\eta$ is implicitly interpreted as TCB in equations of motion of light and planets. However, the physical metric $g_{\alpha\beta}$ differs from $f_{\alpha\beta}$ by the scale factor $a^2(\eta)$, and the time $\eta$ relates to TCB as a Taylor series that can be obtained after expanding $a(\eta)=a_0+\dot a\eta+...,$ in polynomial around the initial epoch $\eta_0$ and defining TCB at the epoch as TCB=$a_0\eta$. Integrating equation $dt=a(\eta)d\eta$ where $t$ is the coordinate time, yields
\be\l{cor3}
t={\rm TCB}+\frac12{\cal H}\cdot{\rm TCB}^2+...\;,
\end{equation}
where ${\cal H}=H/a_0$ and ellipses denote terms of higher order in the Hubble constant ${\cal H}$. The coordinate time $t$ relates to the atomic time TAI (the proper time of observer) by equation, which does not involve the scale factor $a(\eta)$ at the main approximation. It means that in order to incorporate the cosmological expansion to the equations of motion, one must replace TCB to the quadratic form
\be\l{cor4}
{\rm TCB}\longrightarrow {\rm TCB}+\frac12{\cal H}\cdot{\rm TCB}^2\;.
\end{equation}
Distances in the solar system are measured by radio ranging spacecrafts and planets. Equations of light propagation preserve their form if one keeps the speed of light constant and replace coordinates $(\eta,{\bm x})$ to $(t,{\bm\Xi})$, where the spatial coordinates ${\bm \Xi}$ relate to coordinates ${\bm x}$ by equation $d{\bm \Xi}=a(\eta)d{\bm x}$. Because one uses TAI for measuring time, the values of the spatial coordinates in the range measurements are given in terms of the capitalized coordinates ${\bm\Xi}$. Therefore, the ranging measurements are not affected by the time transformation (\ref{cor3}) in contrast to the measurement of the Doppler shift, which deals with time only.
This interpretation may reconcile the "Quadratic Time Augmentation" model of the Pioneer anomaly discussed by Anderson et al (2002) in equation (61) of their paper.
\begin{acknowledgments} This work was supported by the Research Council Grant No. C1669103 and by 2009 Faculty Incentive Grant of the Alumni Organization of the University of Missouri-Columbia.
\end{acknowledgments}

\end{document}